\newcommand{\bz}{{\bf Z}}
\newcommand{\br}{{\bf R}}
\newcommand{\bs}{{\bf S}}

\documentstyle[12pt]{article}

\begin{document}
\begin{titlepage}
\begin{center}

\hfill       PUPT-1659\\
\hfill hep-th/9611042\\

\vskip .9in

{{\Large \bf D-brane field theory on compact spaces}}

\vskip .25in

Washington Taylor IV\\
{\small \sl Department of Physics,} \\
{\small \sl Princeton University,} \\
{\small \sl Princeton, New Jersey 08544, U.S.A.} \\
{\small \tt wati@princeton.edu} \\[0.3cm]
\end{center}
\vskip .25in

\begin{abstract}
We consider Dirichlet $p$-branes in type II string theory on a space
which has been toroidally compactified in $d$ dimensions.  We give an
explicit construction of the field theory description of this system
by putting a countably infinite number of copies of each brane on the
noncompact covering space, and modding out the resulting gauge theory
by $\bz^d$.  The resulting theory is a gauge theory with graded fields
corresponding to strings winding around the torus an arbitrary number
of times.  In accordance with T-duality, this theory is equivalent to
the gauge theory for the dual system of $(d + p)$-branes wrapped
around the compact directions, where the winding number is exchanged
for momentum in the compact direction.
\end{abstract}

\end{titlepage}
\newpage
\renewcommand{\thepage}{\arabic{page}}
\setcounter{page}{1}


D-branes\cite{Polchinski,pcj} are an important new tool for studying
many aspects of string theory, including duality and black hole
physics.  The low energy dynamics of parallel Dirichlet $p$-branes in
noncompact space are neatly described by the dimensional reduction of
10 dimensional supersymmetric gauge theory to $p + 1$
dimensions\cite{Witten-bound}.  A similar field theory description of
D-brane dynamics on general compact spaces which is globally valid has
not yet been given, although a number of important aspects of such
systems have been studied in the context of bound states and black
holes\cite{bsv,Sen,Vafa-gas,Vafa-instantons,sv,dvv}.

Recently, the motion of D-branes on orbifold spaces has been studied
by looking at the motion of multiple copies of each D-brane on the
simply connected covering space and taking a quotient by a discrete
orbifold group\cite{gp,dm,Polchinski-tensors,jm}.  In this note we carry
out an analogous construction for D-branes moving on toroidally
compactified space-times.  We make an infinite number of copies of
each D-brane which we allow to move on a flat noncompact space;
modding out the resulting theory by the lattice $\bz^d$ defining a
$d$-dimensional torus gives a field theory description of D-branes
moving on the quotient space.  The quotient theory thus described is
precisely equivalent as a field theory to the usual gauge theory
description of the T-dual system of D-branes wrapped around the torus.
Thus, although the results of the construction described here agree
with T-duality, the construction itself is independent of duality, and
thus may be useful in studying systems whose T-duals are not already
understood.

%

We now present the quotient construction explicitly in the simplest
case of 0-branes moving on $(\bs^1 \times \br^8) \times\br$.
Thereafter we will discuss briefly the generalization to higher
dimensional branes and higher dimensional compactifications, which is
straightforward.

Let us begin by recalling the gauge theory description of $k$ 0-branes
in type IIA theory in flat space.  The low-energy physics of such a
system is completely described by dimensionally reducing the
10-dimensional super Yang-Mills action to $0 + 1$ dimensions.  In
units where $2 \pi \alpha' = 1$, this gives an action
\begin{equation}
S = \int dt \frac{1}{2g}  \left[ \sum_{i}
{\rm Tr}\; \left( \dot{X}^i -i[A_0, X^i] \right)^2
+ \sum_{i< j}{\rm Tr}\;[X^i, X^j]^2
-i {\rm Tr}\; \bar{\psi} \Gamma^i D_i \psi \right]
\label{eq:0action}
\end{equation}
where $X^i$ and $\psi^i$ are bosonic and fermionic fields written as
$k*k$ matrices.  If we gauge fix $A_0 = 0$, the bosonic part of the
Lagrangian is given by
\begin{equation}
{\cal L} =
\frac{1}{2g} \left[
{\rm Tr}\; \dot{X}^i \dot{X}^i+ \frac{1}{2} {\rm Tr}\;[X^i,
X^j]^2\right].
\label{eq:qm}
\end{equation}
Throughout this paper we will only concern ourselves with the bosonic
part of the action.  The fermionic components can be replaced at any
point using the supersymmetry of the theory.

Since the matrices $X^i$ are hermitian, the classical equations of
motion for the Lagrangian (\ref{eq:qm}) are $[X^i, X^j] = 0$.  When
these equations are satisfied, the matrices are simultaneously
diagonalizable.  In this case, the diagonal elements correspond to
well-defined positions of the 0-branes.  Thus, the moduli space of
classical configurations is just $(\br^9)^k/S_k$ where the quotient by
the symmetric group on $k$ elements arises from the residual gauge
invariance, and expresses the indistinguishability of the particles.
When the matrices do not commute, the 0-branes are smeared out and
cannot be thought of as having classical positions.

We will now compactify one of the spatial dimensions $X^1$ on a circle
$\bs^1$ of radius $R$.  The circle is just the quotient of an infinite
line $\br$ by the discrete group $\Gamma=2 \pi R\bz$.  Thus, we can
describe the physics of 0-branes moving on $(\bs^1\times \br^8)\times
\br$ by making a copy of each 0-brane for each element of $\bz$, and
then imposing the symmetry under $\Gamma$ (for a clear description of
the analogous construction on an orbifold where the group $\Gamma$ is
finite, see \cite{jm}).  If we wish to describe $k$ 0-branes moving on
the compact space, then the usual index $i$ labeling the branes will
be replaced by a pair of indices $j, n$ where $1 \leq j \leq k$ and
$n\in\bz$.  For notational convenience we will write the resulting
infinite dimensional matrices $X^i$ in terms of $k*k$ blocks
$X^i_{mn}$ which satisfy $X_{mn} = X_{nm}^{\dagger}$.  The Lagrangian
for the infinite system of particles moving on the covering space now
reads\footnote{Note: formally, this Lagrangian should be divided by
the (infinite) order of the discrete group $\Gamma$.  However, this
factor will later be multiplied back in when we move to a system of
reduced variables, so we simply drop this factor in all formulae.}
\begin{equation}
{\cal L} =\frac{1}{2g}  \left[
{\rm Tr} \dot{X}_{mn}^i \dot{X}_{nm}^i
+ \frac{1}{2} {\rm Tr}\;(X^i_{mq}X^j_{qn}-X^j_{mq} X^i_{qn})
(X^i_{nr}X^j_{rm}-X^j_{nr} X^i_{rm}) \right]
\end{equation}

Now, in order to describe the physics of the particles moving on the
quotient space, we impose the symmetry under $\Gamma$, which gives the
constraints
\begin{eqnarray*}
X^i_{mn} & = & X^i_{(m-1)(n-1)},\;\;\;\;\; i > 1\\
X^1_{mn} & = & X^1_{(m-1)(n-1)},\;\;\;\;\;  m\neq n\\
X^1_{nn} & = & 2 \pi R I +X^i_{(n-1)(n-1)}
\end{eqnarray*}
The term proportional to the identity matrix in the last line
expresses the condition that the $n$th copy of each 0-brane is
displaced by a distance $2 \pi R n$ from the $0$th copy in the
compactified direction.  As a result of these constraints, all the
dynamical information in the system is contained in the matrices
\begin{equation}
X^i_{n} = X^i_{0n}
\end{equation}
where $(X^i_n)^{\dagger} = X^i_{-n}$.
We can now rewrite the action in terms of these reduced variables.
This is a straightforward calculation; the only subtlety appears in
the term in the potential corresponding to ${\rm Tr}\;[X^1, X^j]^2$.
Expressed in terms of the reduced variables, this term in
the potential becomes
\begin{eqnarray}
\lefteqn{\frac{1}{2g}\left[  \sum_{n = -\infty}^{\infty}
   \frac{(2 \pi Rn)^2}{2}  
{\rm Tr}\; X^j_n X^j_{-n}
 - \sum_{m + n + p = 0}2 \pi R(n-p) X^1_m X^j_n X^j_p
 \right.}\label{eq:1terms}
\\
& &\left.\hspace*{1.4in}-\sum_{k + l + m + n = 0}
 {\rm Tr}\; (X^1_n X^j_m X^1_k X^j_l-X^1_n X^1_m X^j_k X^j_l) \right]
\nonumber
\end{eqnarray}
Writing 
\begin{equation}
S_{n}^{j} = \sum_{q} \left( X^1_{q} X^j_{n-q}-X^j_{q}
X^1_{n-q}\right)
-2 \pi Rn X^j_{n}
= \sum_{q} \left( [X^1_{q}, X^j_{n-q}]\right)-2 \pi R n X^j_{n}
\end{equation}
and
\begin{equation}
T_n^{jk} =\sum_{q}  [X^j_{q}, X^k_{n-q}],
\end{equation}
we can rewrite the Lagrangian in reduced variables as
\begin{equation}
{\cal L} = \frac{1}{2g}  \left[
\sum_{i = 1}^{9} {\rm Tr} \dot{X}_{n}^i \dot{X}_{-n}^i
- \sum_{j = 2}^{9}{\rm Tr}\; S_n^{j}  (S_n^{j})^{\dagger}
- \frac{1}{2}
\sum_{j, k = 2}^{9}{\rm Tr}\; T_n^{jk}  (T_n^{jk})^{\dagger} \right]
\label{eq:0Lagrangian}
\end{equation}
This gives a description of 0-branes moving on the compactified space
in terms of quantum mechanics on an infinite dimensional space.

The classical potential of this theory is minimized when
\begin{eqnarray}
X^i_n & = &  0,\;\;\;\;\; n \neq 0 \label{eq:minimum}\\
{}[X^i_0, X^j_0] & = &  0 \nonumber
\end{eqnarray}
Under these conditions, the hermitian matrices $X^i_0$ are
simultaneously diagonalizable, and have eigenvalues corresponding to
definite 0-brane positions.  To check that the action corresponds to
what we would expect of this physical interpretation, we can expand
around such a classical configuration and compute the masses of the
fields corresponding to strings stretched between distinct 0-branes.
For example, the mass of the field $X^j_n$ corresponding to a string
stretching from a D-brane at position $x^1$ to a D-brane at $y^1$
(with all other coordinates equal) is correctly proportional to
$y^1-x^1-2 \pi R n$, due to the extra $n$ times that the string wraps
around the compact direction.  The other interaction terms in
(\ref{eq:1terms}) similarly correspond to tree-level open string
interactions.

According to T-duality, the system we have just described should be
equivalent to the T-dual field theory of $k$ Dirichlet 1-branes
wrapped around a compact $S^1$ of radius $R' =1/(2 \pi R)$.  In fact, we
can demonstrate this equivalence explicitly by writing a Fourier mode
expansion of the dual theory.  The 1-brane theory is the dimensional
reduction to $1 + 1$ dimensions of 10D SUSY Yang-Mills.  In this
theory, there  is a gauge field $A^{i}, i = 0, 1$ and there are matter
fields $Y^i,  2 \leq i \leq 9$.  In the gauge $A^0 = 0$, the bosonic
part of the action is
\begin{equation}
S = \int dt\frac{dx}{2 \pi R'} 
\frac{1}{2g}\left[   {\rm Tr}\;\dot{Y}^i \dot{Y}^i +
{\rm Tr}\;\dot{A}^1 \dot{A}^1  -
{\rm Tr}\; (\partial_1Y^i -i[A^1, Y^i])^2
 +\frac{1}{2}{\rm Tr}\;[Y^i, Y^j]^2
\right]
\label{eq:1action}
\end{equation}
Writing
\begin{eqnarray*}
A^1 & = &  \sum_{n} e^{inx/R'}  X^1_n\\
Y^i & = &  \sum_{n} e^{inx/R'}  X^i_n
\end{eqnarray*}
and explicitly performing the integral over $x$, we find that
(\ref{eq:1action}) precisely reproduces the Lagrangian
(\ref{eq:0Lagrangian}).

We now make several observations regarding the T-duality between these
theories.  Firstly, this equivalence of field theories gives an
explicit description of how D-brane configurations corresponding to
noncommuting position matrices can be directly converted to gauge
field configurations in the dual theory.  This description can be
compared with other discussions of T-duality in the excitation spectra
of these systems which have been carried out in a perturbative
context, expanding around a fixed classical configuration associated
with definite D-brane positions (see for example\cite{dm,ah}).
Secondly, note that there is a residual symmetry in each of these
theories.  In the 0-brane theory the residual symmetry corresponds to
the fact that our choice of which copy of each brane we chose as the
0th reference copy was arbitrary.  When the matrices are noncommuting
the expression of this residual symmetry is somewhat complicated.
However, when the matrices do commute, the symmetry simply corresponds
to the equivalence of the classical 0-brane configuration under a
shift of each 0-brane's coordinate by an independent multiple of $2
\pi R$ in the compactified direction.  In the dual 1-brane theory,
this symmetry just corresponds to the symmetry under global gauge
transformations which are not connected to the identity.  Because the
space of matrices which satisfy (\ref{eq:minimum}) and thus minimize
the classical potential in the 0-brane theory must be modded out by
this symmetry, the space of classical configurations of the 0-branes
is given as expected by $(S^1 \times\br^8)^k/S_k$.

We have described here explicitly the case of 0-branes moving on a
space which has been compactified in a single direction.  The
generalization to parallel Dirichlet $p$-branes of arbitrary dimension
moving on a space which has been compactified in $d$ directions is
straightforward.  In the general case, the indices of each field
become a $d$-tuple of integers, corresponding to winding/Fourier modes
on the $d$-torus.  The only other substantial difference in the
construction is that the potential terms analogous to (\ref{eq:1terms})
becomes slightly more complicated.  When two fields $X^i$ and $X^j$
both correspond to compactified directions, a term appears in
(\ref{eq:0Lagrangian}) of the form $ (-{\rm Tr}\;  Q_{nm}^{ij}
(Q_{nm}^{ij})^{\dagger})$ where
\begin{equation}
Q_{nm}^{ij} 
= \sum_{q, r} \left( [X^i_{q, r}, X^j_{n-q, m-r}]\right)-
2 \pi R n X^j_{n, m}
-2 \pi R m X^i_{n, m}.
\end{equation}
(All winding indices except the two of interest have been dropped, but
of course generically all fields will have $d$ integer indices.)  In
the T-dual theory, such a term corresponds to a curvature term
$F_{ij}$ in a pair of compactified directions around which the dual
D-branes are wound.

We conclude this note with a brief discussion of several situations in
which the construction described here may be relevant.  In the
situation described here of parallel D-branes of the same dimension,
the quotient field theory we constructed was precisely equivalent to
the known field theory of the T-dual system.  However, there are
situations where this construction might be applied where no simple
T-dual field theory is known.  An example of such a situation is a
system of interacting $p$-branes and $p + 4$-branes where the latter
objects are wrapped around a 4-dimensional compact manifold.  Systems
of interacting branes of this type have been used in recent studies of
black holes in string theory\cite{sv,cm}.  The spectrum of BPS bound
states of these systems can be understood from duality
arguments\cite{Sen} or by quantizing the five-brane\cite{dvv}, and can
be counted explicitly from the D-brane picture
asymptotically\cite{sv,cm} and exactly in special
cases\cite{Vafa-gas,bsv2}.  Furthermore, it has been suggested that
the moduli space of classical configurations of such a system can be
identified with the moduli space of instantons on the compact
space\cite{Vafa-instantons}.  This suggestion agrees with the
observation\cite{Witten-small,Douglas} that instantons in the $p +
4$-brane world-sheet theory carry $p$-brane charge.  In
\cite{bsv} the matter fields in such a theory corresponding to ND
strings connecting the $p$- and $p + 4$-branes were described.
However, because there is no general description of a global field
theory for the system of interacting D-branes on a compact space, it
has been difficult to prove this conjecture by explicitly constructing
the moduli space.  The situation is better in noncompact space, where
it has been shown that if the gauge fields on the $p + 4$ branes are
fixed, the moduli space of vacua has an explicit construction as a
hyperk\"ahler quotient\cite{Witten-small,dm}.  This hyperk\"ahler
quotient construction is precisely equivalent to the ADHM construction
of instantons on $\br^4$, showing that in this case the moduli space
of instantons arises directly from the D-brane field theory.  There
are several obstacles to repeating this argument in the compact case,
including the fact that there is no finite dimensional analogue to the
ADHM construction for instantons on compact 4-manifolds and the
absence of a global field theory formulation of the intersecting
D-brane system.  The approach described here, which provides a global
description of a D-brane system on a compact space as a quotient of a
noncompact system, may provide a tool for better understanding systems
of interacting branes of this type and may shed light on the
connection between such systems and instanton moduli space.  Work in
this direction is in progress.

Another possible application of the method described here is to
0-brane quantum mechanics on compact spaces.  Recently, it was
suggested that the 0-brane matrix quantum mechanics described by the
action (\ref{eq:0action}) is more than just a low energy description
of 0-branes in type IIA string theory, and actually contains within it
at least a large part, if not all, of the physics of 11-dimensional
M-theory\cite{bfss}.  In particular, the fact that this matrix quantum
mechanics contains naturally within it the 11-dimensional
supermembrane\cite{dhn,dln} gives strong support to this conjecture.
It is clearly important to understand how membranes in this theory
would behave upon compactification.  Perhaps the approach outlined
in this note will be helpful in elucidating this issue.

I would like to thank Mike Douglas, Aki Hashimoto, Juan Maldacena,
Samir Mathur, Greg Moore, and Herman Verlinde for helpful
conversations.  This work was supported by the National Science
Foundation (NSF) under contract PHY96-00258.


\begin{thebibliography}{999}
\parindent=.6em
\bibitem{Polchinski} J.\ Polchinski, {\em  Phys.\ Rev.\ Lett.} 
{\bf 75}  (1995) 4724, hep-th/9510017.
\bibitem{pcj} J.\ Polchinski, S.\ Chaudhuri and C.\ V.\ Johnson, {\em
Notes on D-branes}, hep-th/9602052.
\bibitem{Witten-bound} E.\ Witten, {\em Nucl. Phys.} {\bf  B460}
(1996) 335, hep-th/9510135.
\bibitem{bsv} M.\ Bershadsky, C.\ Vafa and V.\ Sadov, {\em
Nucl. Phys.} {\bf B463} (1996) 398, hep-th/9510225.
\bibitem{Sen}  A.\ Sen, {\em  Phys.\ Rev.} {\bf  D53} (1996) 2874,
hep-th/9511026. 
\bibitem{Vafa-gas} C.\ Vafa, {\em Nucl. Phys.} {\bf B463},
(1996) 415, hep-th/9511088.
\bibitem{Vafa-instantons}  C.\ Vafa, {\em Nucl. Phys.} {\bf B463}
(1996) 435, hep-th/9512078.
\bibitem{sv} A.\ Strominger and C.\ Vafa, {\em  Phys.\ Lett.} {\bf
B379} (1996) 99, hep-th/9601029.
\bibitem{dvv} R.\ Dijkgraaf, E.\ Verlinde and H.\ Verlinde,
{\em BPS quantization of the five-brane}, hep-th/9604055;
{\em  Counting dyons in $N = 4$ string theory}, hep-th/9607026.
\bibitem{gp}  E.\ Gimon and J.\ Polchinski, {\em  Phys.\ Rev.} {\bf
D54} (1996) 1667, hep-th/9601038. 
\bibitem{dm} M.\ Douglas and G.\ Moore, {\em  D-branes, quivers, and
ALE instantons}, hep-th/9603167.
\bibitem{Polchinski-tensors} J.\ Polchinski, {\em  Tensors from K3
orientifolds}, hep-th/9606165.
\bibitem{jm} C.\ Johnson and R.\ Myers, {\em  Aspects of type IIB
theory on ALE spaces}, hep-th/9610140.
\bibitem{dm2} S.\ Das and S.\ Mathur,
{\em  Interactions involving D-branes}, hep-th/9607149.
\bibitem{ah} A.\ Hashimoto, {\em  Perturbative dynamics of fractional
strings on multiply wound D-strings}, hep-th/9610250.
\bibitem{cm}  C.\ Callan and J.\ Maldacena, {\em Nucl. Phys.} {\bf
B472} (1996) 591, hep-th/9602043.
\bibitem{bsv2} M.\ Bershadsky, C.\ Vafa and V.\ Sadov, {\em
Nucl. Phys.} {\bf B463} (1996) 420, hep-th/9511222.
\bibitem{Witten-small} E.\ Witten, {\em Nucl. Phys.} {\bf  B460}
(1996) 541, hep-th/9511030.
\bibitem{Douglas}  M.\ Douglas, {\em  Branes within branes},
hep-th/9512077; {\em Gauge fields and D-branes}, hep-th/9604198.
\bibitem{bfss} T.\ Banks, W.\ Fischler, S.\ H.\ Shenker and L.\
Susskind, {\em  M theory as a matrix model: a conjecture}, hep-th/9610043.
\bibitem{dhn} B.\ de Wit, J.\ Hoppe and H.\ Nicolai, {\em Nucl. Phys.}
{\bf  B305}[FS 23] (1988) 545.
\bibitem{dln} B.\ de Wit, M.\ Luscher and H.\ Nicolai, {\em
Nucl. Phys.} {\bf  B320} (1989) 135.
  
\end{thebibliography}
\end{document}